\documentclass[%
 reprint,
 superscriptaddress,
 showpacs,preprintnumbers,
 amsmath,amssymb,
 aps,
pra,
 floatfix,
]{revtex4-1}

\usepackage{graphicx}% Include figure files
\usepackage{dcolumn}% Align table columns on decimal point
\usepackage{bm}% bold math
\usepackage[colorlinks,breaklinks,
 citecolor=red,linkcolor=red,urlcolor=blue
]{hyperref}% add hypertext capabilities
\begin{document}
\bibliographystyle{apsrev4-1}

\title{Cold collision and high-resolution spectroscopy of buffer gas cooled BaF Molecule}%

\author{Wenhao Bu}%
\author{Tao Chen}
\author{Guitao Lv}
\author{Bo Yan}%
 \email{yanbohang@zju.edu.cn}
\affiliation{%
 Department of Physics, Zhejiang University, Hangzhou, China, 310027
}%

%\collaboration{}%\noaffiliation

\date{\today}

\begin{abstract}
We reported a detailed experimental study of the cold collision of Barium monofluoride (BaF) with buffer gas and the high-resolution spectroscopy relevant with direct laser cooling. BaF molecules are efficiently produced with laser ablation and buffer-gas cooled in a cryogenic apparatus. The laser cooling relevant transition $|X^2\Sigma, v=0, N=1\rangle$ to $|A^2\Pi, v'=0, J'=1/2\rangle$ is identified. The collision cross section with buffer gas is measured to be $1.4(7)\times10^{-14} cm^{-2}$, which is very suitable for buffer gas cooling. Both rotational and vibrational temperatures are effectively cooled, and large mount of molecules are quenched into the desired states. Our study provides an important benchmark for further laser cooling of BaF molecule.
\end{abstract}
%\pacs{34.20.Cf, 33.20.Vq, 31.15.am, 37.10.Mn}
\maketitle

%\tableofcontents

\section{\label{section1}Introduction}
Cold molecules have plenties of important applications in fields ranging from quantum chemistry \cite{Ni2008,Ni2010,Ospelkaus2010,Quemener2012}, quantum simulation \cite{Micheli2006,Yan2013,Gadway2016}, to precise measurement \cite{Hudson2011,Baron2014}, and so on. Over the last 20 years, the development of cold molecule physics has been pushed one after another, from traditional method by taming molecules with electric and magnetic field, such as Stark decelerator \cite{Bethlem1999,Bochinski2003,VanDeMeerakker2008} and Zeeman decelerator \cite{Narevicius2008,Raizen2009}, to the laser cooling technique. Direct laser cooling, which is believed to be able to bridge the "$\mu \text{K}$" gap \cite{Carr2009}, is relatively a new direction and has achieved great progress in recent years. Lots of diatomic molecules have been laser cooled\cite{Shuman2010,Norrgard2016,Hummon2013,Yeo2015,Zhelyazkova2014,Hemmerling2016} or under explored \cite{Tarallo2016,Xu2016,Smallman2014}, and even polyatomic ones have recieved signifcant attentions\cite{Zeppenfeld2012,Kozyryev2015,Prehn2016}. Among these molecules, BaF is a good candidate for laser cooling \cite{Chentao}, the Franck-Condon factors are feasible for quasi-cycling transition. The cooling and repumping transitions are around $900 \text{nm}$, in the good regime of the diode laser, and the rotational constant is about $6.5~\text{GHz}$ \cite{Ryzlewicz1980}, which can be easily dressed with microwaves.

Most of the molecules which can be laser cooled are highly reactive, and they are nonexistent in nature, thus must be produced in laboratory from gases or solid precursors. Molecules created in this way are usually very hot, the temperature is as high as $\sim 1000~\text{K}$. For laser cooling considerations, pre-cooling such hot molecules is of great importance to obtain a stable molecular source. Buffer-gas cooling is a very good method for such purpose, and has been developed as an effective process to produce a molecular beam \cite{Hutzler2012}. For buffer gas cooling, both the elastic and inelastic collisions with cold buffer gas dissipate the energy of molecules, making all the degrees of freedom of molecules be cooled \cite{Skoff2011,Maxwell2005,Barry2011,Bulleid2013,Kozyryev2015}. Not only the translational velocity can be efficiently slowed down, but also the molecules are largely enriched in the lowest vibrational level and a few lower-lying rotational levels, making the science states used in laser cooling scheme largely populated.

Another concerned issue for further BaF laser cooling experiments is the identification of the cooling and repumping laser transitions. For the current in-cell stage experiment, direct measurement on the absorption spectroscopy is a simple and practical means to acquire the collision properties and the required transition information. Meanwhile, the temperature of the BaF molecule can be resolved to check the efficiency of the buffer gas cooling.

In this paper, we demonstrate the effective creation of BaF molecules from laser ablation, and buffer gas cooling with the 4 K helium gas. The cold collision cross section and the high-resolution spectroscopy relevant with laser cooling are measured. In Sec.\ref{section2}, we describe the experimental setup, including the 4 K cryogenic apparatus, the BaF$_2$ target making and the absorption measurement. Section \ref{section3} gives the analyses on the absorption signal. We report the effective production of BaF molecules by laser ablation, and the measurement of the high-resolution spectroscopy helps to identify the main laser cooling transition. Furthermore, the collisional cross section between BaF and helium is measured and the efficient rotational and vibrational cooling are observed. In the final section, we conclude and summarize our results.

\section{\label{section2} Experiment}
\begin{figure}[]
\includegraphics[width= 0.5\textwidth]{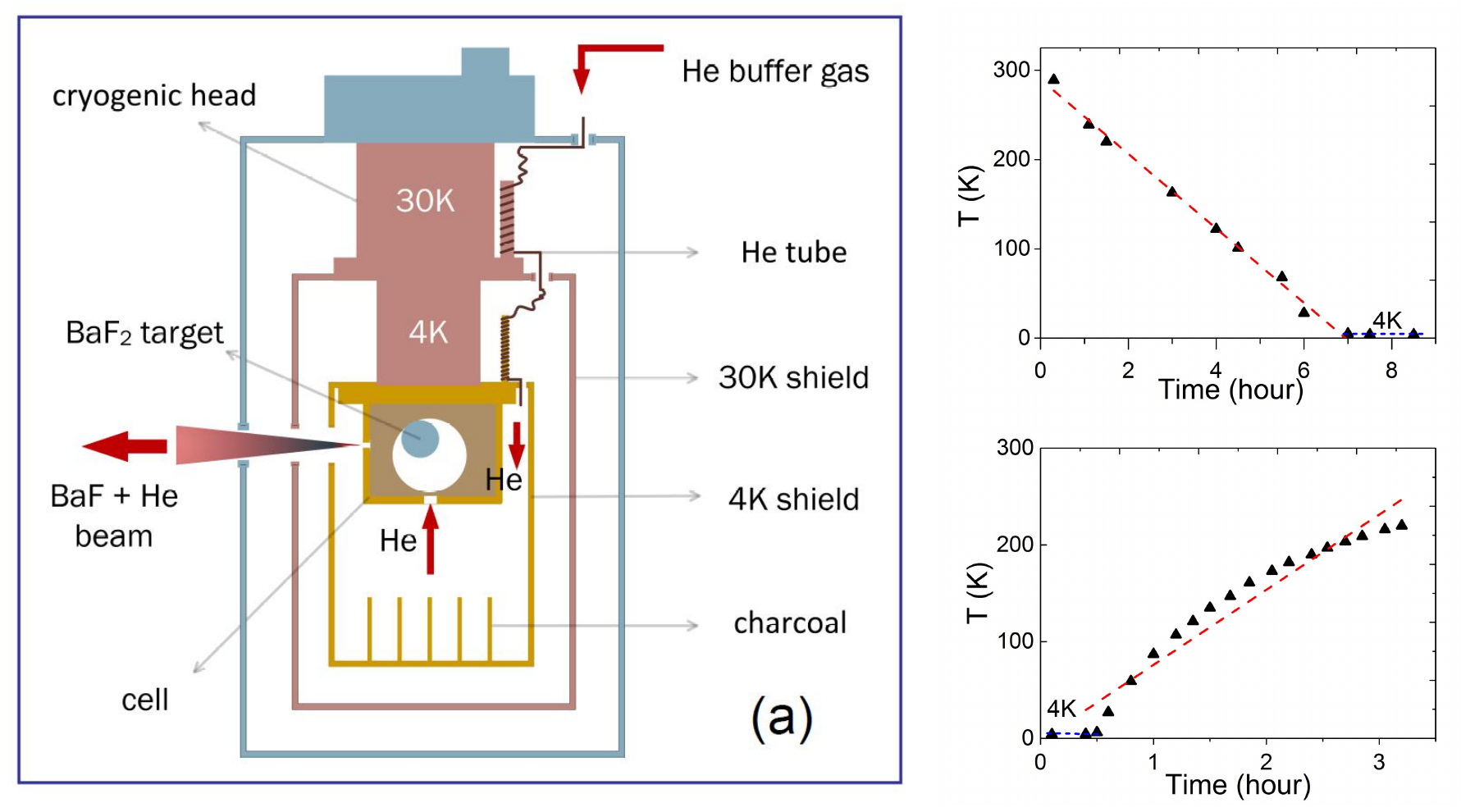}
\caption{\label{figure1}(Color online) Schematic experimental setup. (a) The cryogenic apparatus. The whole system is a aluminum vacuum chamber, the $30$ K and 4 K shielding layers are attached to the first stage and the second stage coolers of a pulse tube refrigerator respectively. The buffer gas is pre-cooled by the 30 K and 4 K heat sinks before sending into the cell. Charcoal is used to help pump helium gas at 4K. (b) The cooling process from room temperate to 4 K. It takes roughly $8$ hours. (c) The warming process from 4 K to room temperature.}
\end{figure}

The BaF molecules are created with laser ablation followed by buffer gas cooling. Figure \ref{figure1}(a) shows the schematic cryogenic apparatus, including a pulse tube refrigerator (Sumitomo, SRP-082B-F70H), which can achieve a temperature of 30 K at the first stage, and 4 K at the second stage. The vacuum chamber contains three layers of shielding. The outer one is an aluminum vacuum chamber, which provides an vacuum of better than $10^{-6}$ Pa. The vacuum is maintained by a $300~\text{L/s}$ turbo pump. The second layer is the 30 K shielding, which is attached to the first stage cryogenic head of the pulse tube refrigerator. The center one is the 4 K shielding, and attached to the second stage cryogenic head. In order to minimize the black-body radiation effect from the environment, the 30 K and 4 K shielding layers are coated with gold. We also put a few copper layers attached with activated charcoal in the 4 K shielding box to enhance the pump speed of the helium gas. The helium buffer gas is injected into the chamber through a steel-copper tube, and is pre-cooled to 30 K and 4 K by the heat sinks before running into the cell via the aperture on the bottom side of the cell; see the red arrow in Fig.\ref{figure1}(a). The flow rate of the helium buffer gas is controlled and can be tuned from 1 to 20 sccm (standard cubic centimeters per minute) by a flowmeter. In order to avoid the tube being blocked due to the frozen of other contaminant gas, the purity of helium gas is chosen to be $99.999\%$. The science cell is made of a piece of cubic copper $(L=45~\text{mm})$, with two cross holes $(r=10~\text{mm})$. It takes roughly $8$ hours to cool the cell from room temperature to 4 K and a similar time back to 300 K, as shown in Fig.\ref{figure1}(b) and (c).

Due to the naturally unstable property of BaF, we have to produce it artificially in the laboratory. Generally, two methods are used to create such diatomic molecules. One is laser ablation, and another is chemical reaction \cite{Tu2009,Zhelyazkova2014}. The later method can produce molecules with higher flux , while the first one is much simpler to operate. We use laser ablation in our experiment. The ablation laser is a 532 nm pulse laser (Lapa-80, Beamtech Optronics), with a pulse duration $12~\text{ns}$, the repetition rate can be tuned from $1-20\text{Hz}$, and the highest output power is $50~\text{mJ}$.

The target in Fig.\ref{figure1}(a) is made from BaF$_2$ powder, which shows rather weak viscidity. Consequently, we had to mix $\sim 5\%$ CaF$_2$ powder to increase the viscidity of the samples, and it indeed works very well. The mixed BaF$_2$ and CaF$_2$ powder was compressed with a pressure of $\sim 500~\text{MPa}$ for 10 minutes to produce a pill shaped sample. Then the pills were heated up to $1000~^\circ\text{C}$ and baked for 3 hours to make them stiff.

The laser ablation and absorption spectroscopy are performed in the science cell, as schematically illustrated in Fig.\ref{ablation}(a). The ablation 532 nm pulse laser is focused to a $1/e^2$ diameter of $0.5~\text{mm}$ on the BaF$_2$ target by a $f = 400~\text{mm}$ lens. An absorption spectroscopy scheme is setup to detect and monitor the molecule creation and collision dynamics. The probe beam passes through the cell and is detected by a photodetector (PD). We use a transfer cavity locked diode laser to ensure the frequency stability \cite{Bu2016}. Once the $860~\text{nm}$ laser hits the molecular transitions, absorption happens and a dip appears at the time trace of the PD signal.

\section{\label{section3} Results and Discussions}

\subsection{\label{subsectionA} Absorption signal from single shot}
\begin{figure}[]
\includegraphics[width= 0.5\textwidth]{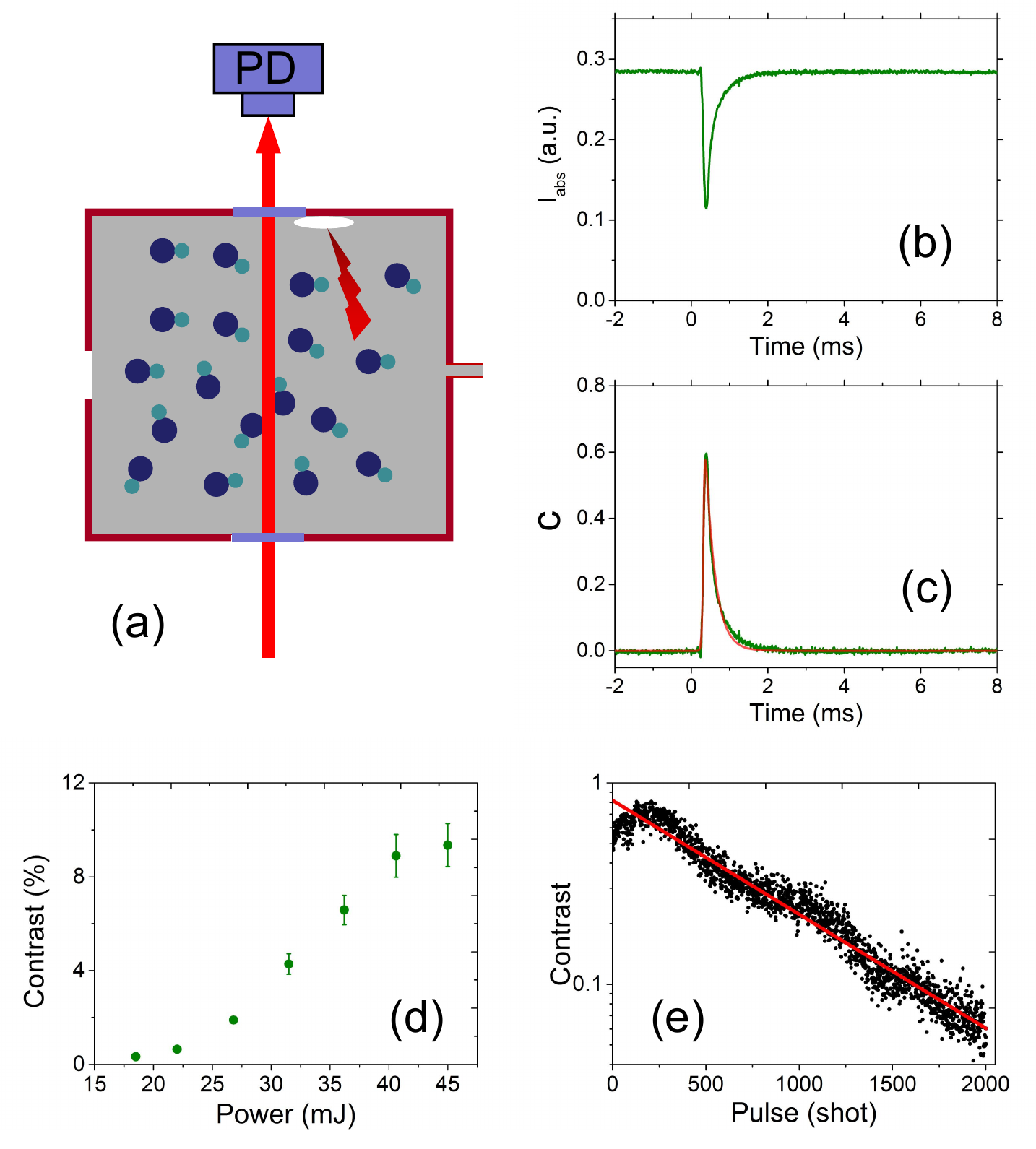}
\caption{\label{ablation}(Color online) (a) The experimental setup for absorption spectroscopy measurement. The BaF molecules are produced by laser ablation in the buffer gas cell. (b) The typical absorption signal of molecules with laser ablation. The pulse laser fires at $t=0~\text{ms}$, the absorption signal shows up and then decays. (c) The normalized signal. The red line is the fitting function with formula (\ref{eq3}). (d) The ablation laser power dependence of molecule production. The pulse laser is $532 ~\text{nm}$, 2 Hz, $12~\text{ns}$, $1/e^2$ diameter is $0.5~\text{mm}$. (e) The decay of laser ablation at one point. When the target are hit by laser more and more at one point, the produced molecule number decay. The decay constant is about few hundreds shots. All the data are measured with the He buffer gas rate of 5 sccm. }
\end{figure}

The typical absorption signal from single shot of the ablation laser is shown in Fig.\ref{ablation}(b). In order to extract the information behind the signal, we first normalized it to the absorption fraction
\begin{equation}
c=1-\frac{I_\text{abs}}{\text{Max}(I_\text{abs})},
\end{equation}
as shown in Fig. \ref{ablation}(c). Then we construct a two-step model. Firstly, the molecules creation process is modeled by a logistic function
\begin{equation}
y=\frac{A}{1+e^{-(t-t_0)/\tau_1}},
\end{equation}
where $t_0$ means the delay after the pulse laser fired, and it depends on the distance between the target and the probe beam. $\tau_1$ evaluates how fast the creation process happens.

The second step is the decay process caused by the diffusion of molecules in the buffer gas cell, including the colliding with the cell walls and inelastic collisions with the buffer gas. Over all, we can model it with an exponential decay, so the two-step model gives a fitting function,
\begin{equation}
y=\frac{A}{1+e^{-(t-t_0)/\tau_1}}\cdot e^{-(t-t_0)/\tau_2}. \label{eq3}
\end{equation}
Here $\tau_2$ describes how fast the signal decays. This two-step model fits quite well with the experimental data as shown in Fig.\ref{ablation}(c).

From the fitting function (\ref{eq3}), we define the contrast, \emph{i.e.}, the maximum absorption fraction, to reflect the molecule number from laser ablation. Figure \ref{ablation}(d) shows the ablation laser power dependence of the molecule creation. The contrast starts to saturate around $40~\text{mJ}$, and this point is set for the daily operation. Another issue is the decay of signal when we continually hit one point on the BaF$_2$ target with the ablation laser. Figure \ref{ablation}(e) shows a typical data for such measurement. The lifetime for laser ablation at one position is about a few hundreds shots. In order to maintain good molecule creation condition, we need to change the fire point by adjusting the direction of the ablation laser from time to time.

\subsection{\label{subsectionB}Collisional cross section with buffer gas}

\begin{figure}[]
\includegraphics[width= 0.4\textwidth]{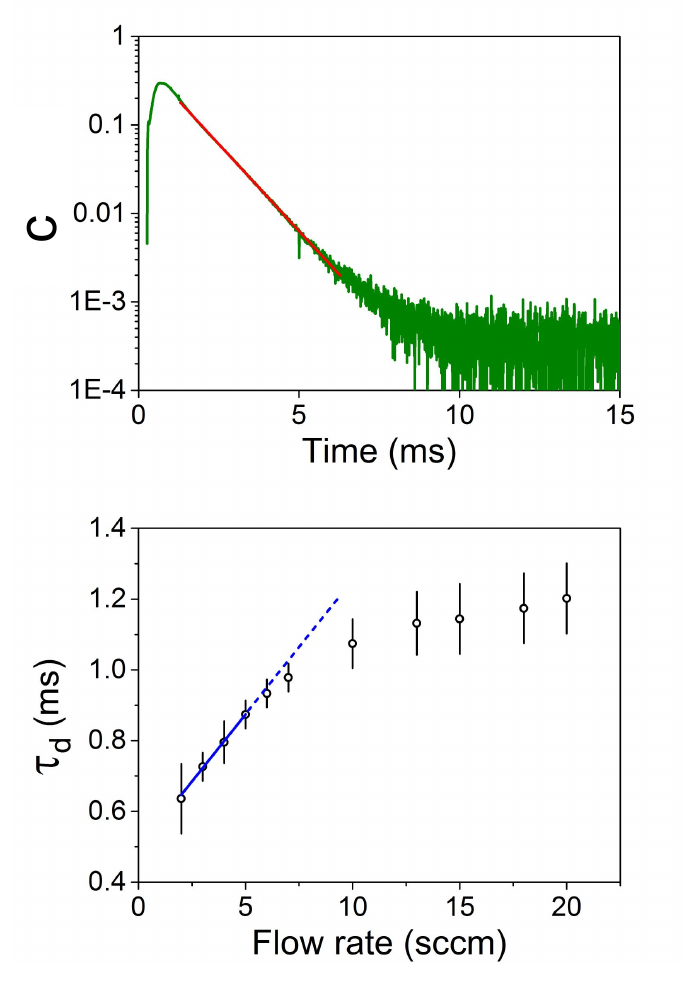}
\caption{\label{decay_time}(Color online) The cold collision of BaF with buffer gas. (a) The absorption fraction decay curve with log scale. Long time after ablation, the high mode of molecule diffusion decays away. Then the curve can be fitted with a single exponential decay. The red line is the exponential decay fitting. (b) The decay time constant versus the flow rate. Higher flow rate means higher helium density, and thus lead to slower diffusion. The helium density linearly follows the flow rate at the diffusive limit when the flow rate is small. We fit data linearly at the beginning to get the slope to calculate the collision cross section.}
\end{figure}

As shown in Fig.\ref{ablation}(b),  once the molecules are created by the ablation laser, it decays. The relaxations are mainly caused by the diffusion to the cell walls and the leakage via the aperture to form a molecular beam. Figure \ref{decay_time}(a) plots the absorption fraction signal on logarithmic scale, where the helium flow rate is 5 sccm. Under this condition, the flow rate is relatively low, the diffusion to the cell walls dominates in molecule loss, and consequently the in-cell lifetime of the BaF molecule is proportional to the helium density in the cell \cite{Patterson2007}. Within the first 2 ms, the high-order diffusion mode rapidly decays. However, from 2 ms to 6 ms, the logarithmic signal can be fitted well with a line, which in fact reflects the in-cell colliding dynamics. The decay time constant $\tau_\text{d}$ depends on the collision cross section with a simple formula \cite{Kozyryev2015},
\begin{equation}
\tau_\text{d}=\frac{n_\text{He}\sigma_\text{d}}{\bar{v}G},
\end{equation}
where $n_\text{He}$ is the helium density, $\sigma_\text{d}$ is the collision cross section between molecule and the buffer gas, $\bar{v}=(8k_\text{B}T/\pi\mu)^{1/2}$ is the average velocity between molecules and the helium gas at temperature $T$, and $\mu$ is the reduced mass of the BaF-He system.

For a cylindrical tube,
\begin{equation}
G=\frac{3\pi}{32}(\frac{2.405^2}{r^2}+\frac{\pi^2}{L^2}),
\end{equation}
where $r$ and $L$ are the radius and length of the cylindrical tube respectively. In our experiment, we control the flow rate of helium gas into the cell from $2-20$ sccm. The helium density is determined by \cite{Barry2013}
\begin{equation}{\label{eq6}}
n_\text{He}=\frac{\kappa f}{A_\text{aperture}\bar{v}_\text{b}},
\end{equation}
where $A_\text{aperture}$ is the area of the aperture and it is $78~\text{mm}^2$ in our case, $\bar{v}_{b}$ is the mean velocity of the buffer gas. At the full effusive limit (low helium flow rate regime), $\kappa=2\sqrt{\pi}$, and at the supersonic regime, $\kappa$ could be a factor of $2$ lower \cite{Barry2011}.

In Fig.\ref{decay_time}(b), we plot the decay time constant $\tau_\text{d}$ versus the helium flow rate. By fitting it with a linear function at low flow rate regime, the collision cross section between BaF and helium is estimated from formula (\ref{eq6}) at the effusive limit. It is about $1.4(7)\times10^{-14}~\text{cm}^{-2}$. The error mainly comes from uncertainties of the determination on the helium density $n_\text{He}$ and the geometry factor $G$ of the cell.

\begin{figure}[]
\includegraphics[width= 0.35\textwidth]{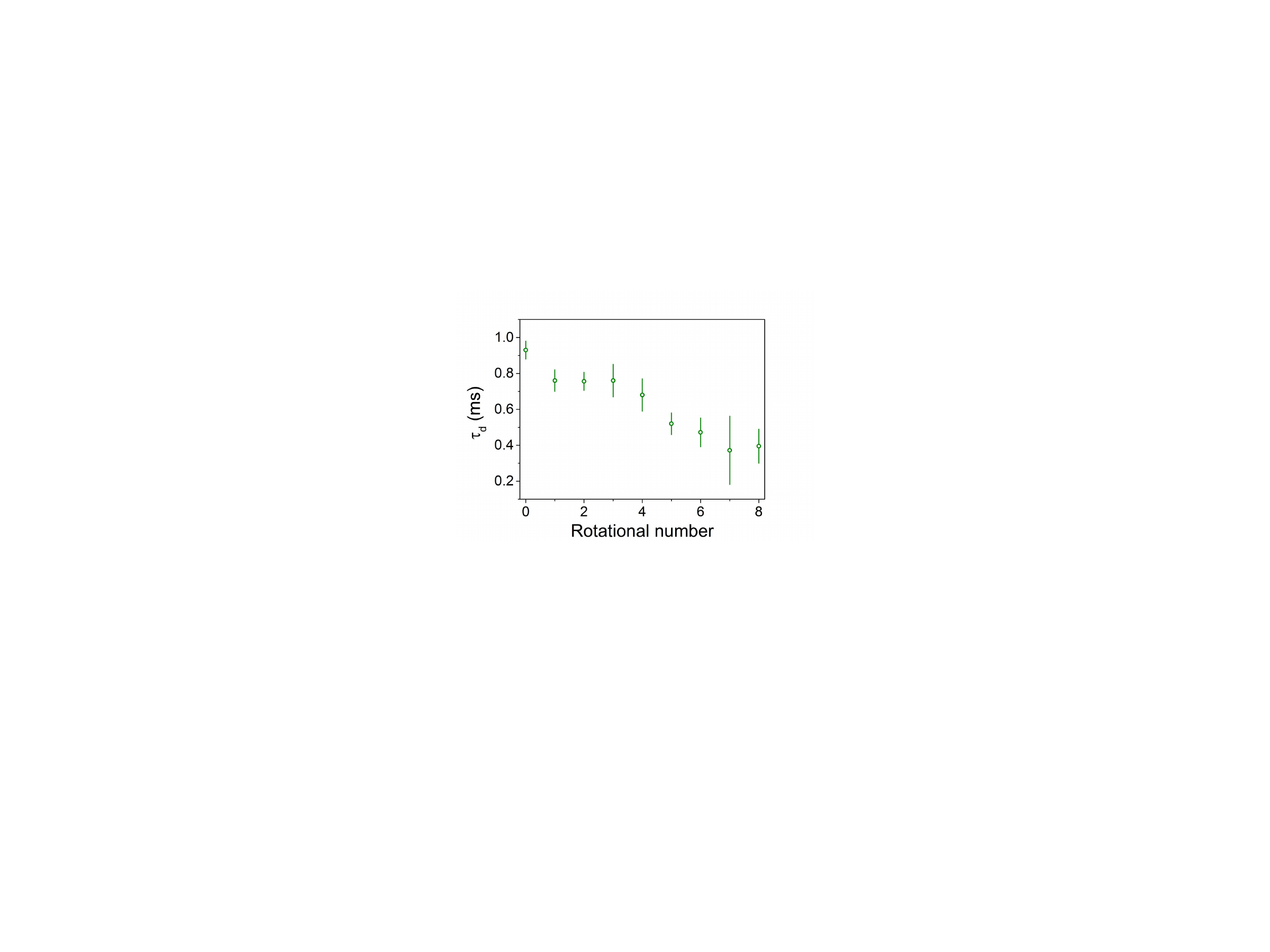}
\caption{\label{rot}(Color online) Rotational number versus decay time constant. The helium flow rate is 5 sccm.}
\end{figure}

We also measured the decay time constants for different rotational states ($|X^2\Sigma, v=0\rangle$). The helium flow rate is fixed to be 5 sccm, and the probe laser is scanned to find the transitions for different rotational states. Then we picked the resonance frequencies and measured the absorption spectroscopies, and extract $\tau_d$. As shown in Fig. \ref{rot}, $\tau_d$ decreases when the rotational number increases. It might be due to the collision cross section of BaF with helium is different for different rotational number at this temperature, as theoretic suggested for other molecules \cite{Gonzalez-Sanchez2008}.

\subsection{\label{subsectionC} High-resolution spectroscopy}

\begin{figure}[]
\includegraphics[width= 0.5\textwidth]{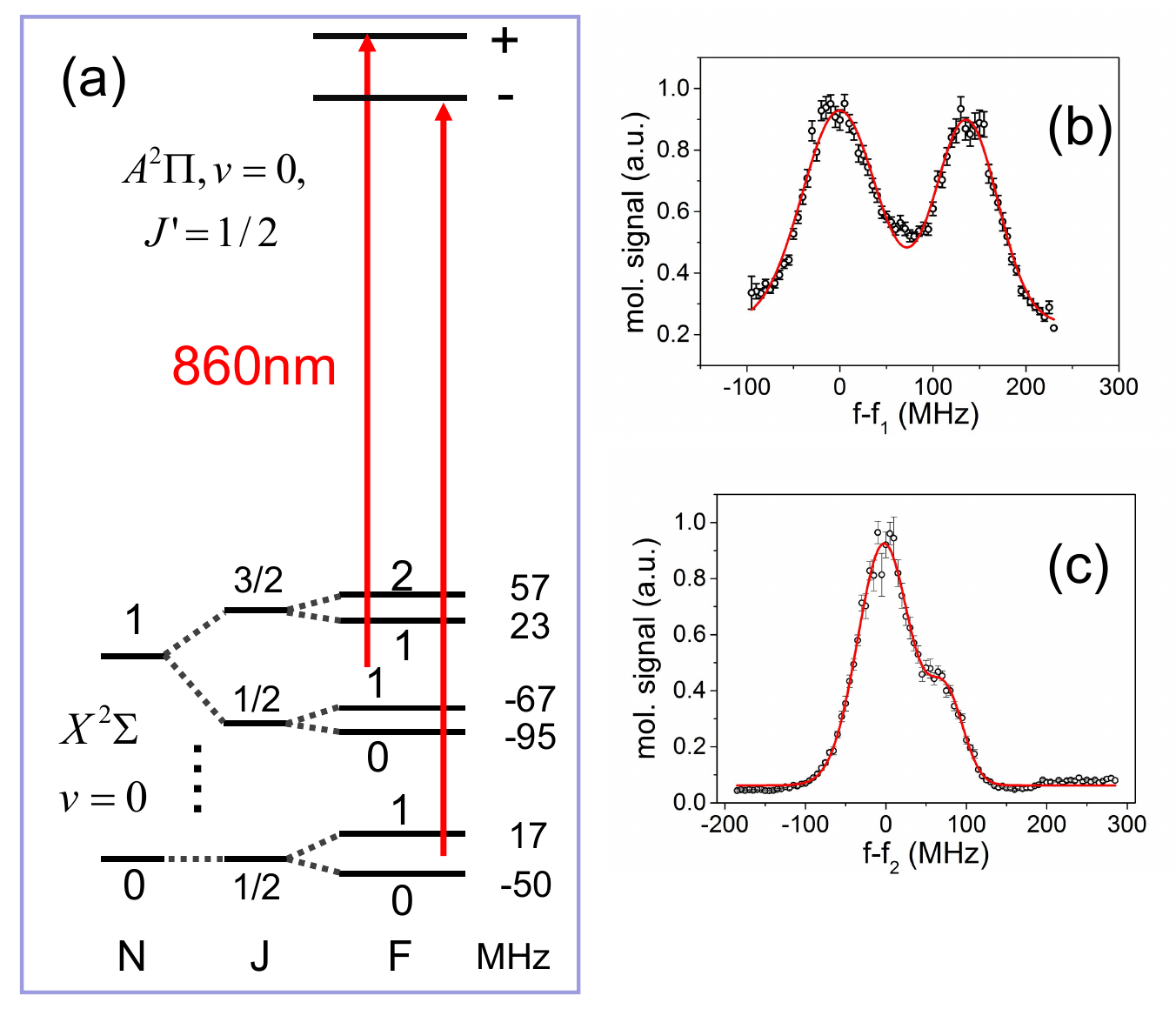}
\caption{\label{hyperfine}(Color online) High-resolution spectroscopy. (a) The main cooling transition of BaF molecules. The hyperfine levels for $N=0$, and $N=1$ are shown. Because of the parity selection rule, the transitions from $|X, N=1\rangle$ to $|A, J'=1/2, +\rangle$ and $|X, N=0\rangle$ to $|A, J'=1/2, -\rangle$ are allowed. (b) The absorption spectroscopy of transition from $|X, N=1\rangle$ to $|A, J'=1/2, +\rangle$, $f_1=11630.0843\ \text{cm}^{-1}$. (c) The absorption spectroscopy of transition from $|X, N=0\rangle$ to $|A, J'=1/2, -\rangle$. $f_2=11630.2595\ \text{cm}^{-1}$. The hyperfine structures are resolved.}
\end{figure}

Laser cooling technique requires the lasers to be stabilized to $1\sim2$ MHz and the transition should be identified to be within a few MHz. High-resolution spectroscopy should be performed to find those transitions. Fortunately, the spectroscopy of BaF free radical has been studied for a long time, and the molecular constants have been determined very accurately \cite{Ernst1986,Guo1995,Berg1998,Ryzlewicz1980,Steimle2011,Chentao}. Figure \ref{hyperfine}(a) shows the main cooling transitions from $|X^2\Sigma, v=0, N=1\rangle$ to $|A^2\Pi, v'=0, J'=1/2, +\rangle$. The hyperfine structure of lower $N=1$ level has splittings round $30$ MHz and $120$ MHz, while the hyperfine splitting of the exciting state $|A^2\Pi\rangle$ is about a few MHz, and is hard to resolve.

In spectroscopy experiments, the transiton $\vert N, J \rangle \to \vert N', J' \rangle$ is labeled by $^{\Delta N}\Delta J (N)$, where $\Delta N = N'-N$, $\Delta J = J'-J$. $\Delta J (\text{or}\ \Delta N) = -1,0,+1$ are corresponding to the $P, Q, R$ branches. Figure \ref{hyperfine}(b) shows the typical in-cell spectrum of the main cooling transition $|X^2\Sigma, v=0, N=1\rangle \to |A^2\Pi, v'=0, J'=1/2, +\rangle$. This transition involves the $P(1)$ and $^PQ(1)$ branches, corresponding to the left and right peaks of the spectrum. The separation between the two peaks is $135(5)\ \text{MHz}$, and it is consistent with the calculated energy difference between the $J=1/2$ and $J=3/2$ fine structrue levels \cite{Chentao}. The full width half maximum is about $80(8)\ \text{MHz}$, much bigger than the Doppler broadening $(40\ \text{MHz})$ for BaF at $T=4$ K, which might be caused by collision with the buffer gas. For the current-state experiment, we can not resolve $F=0,1$ and $F=1,2$ states (the hyperfine structure) with such large broadening.

We have also measured the $Q(0)$ transition from $|X^2\Sigma, v=0, N=0\rangle$ to $|A^2\Pi, v'=0, J'=1/2, -\rangle$, as shown in Fig.\ref{hyperfine}(c). The separation of the two peaks is $77(5)\ \text{MHz}$, which is consistent with the energy gap between $F=0$ and $F=1$ states for $|v=0, N=0\rangle$. The hyperfine structure of the N=0 rotational state is resolved. The absolute value of the wavelength in Fig.\ref{hyperfine} comes from a ultra-high resolution wavemeter (High-Finesse WS-7), with an uncertainty of 60 MHz, which is the limit of our measurement.

\subsection{\label{subsectionD}Rotational and vibrational cooling}
\begin{figure}[t]
\includegraphics[width= 0.45\textwidth]{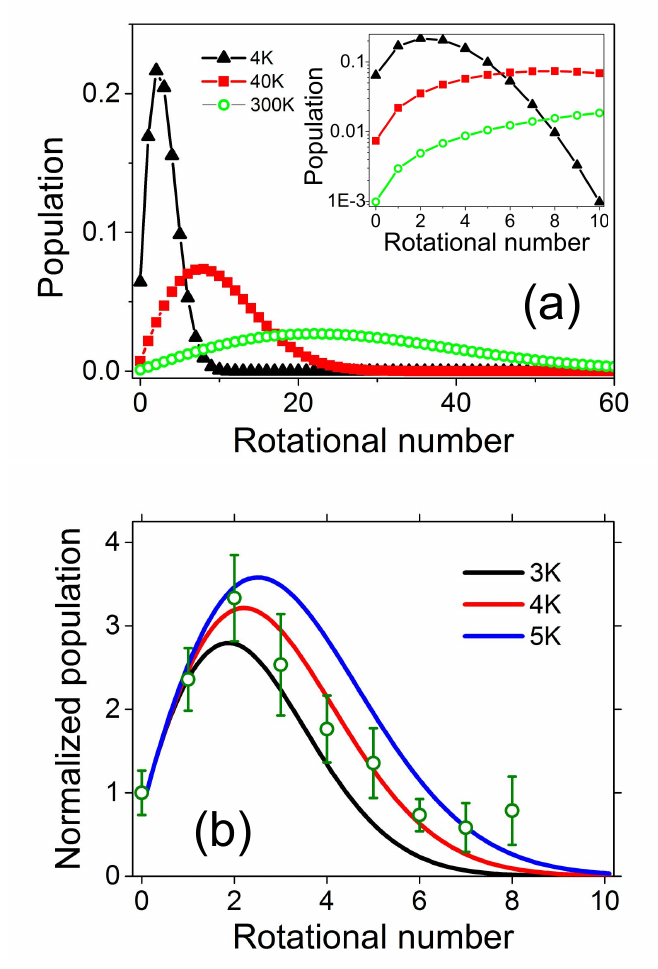}
\caption{\label{rotational}(Color online) The rotational relaxation for buffer gas cooling. (a) The theoretic calculation of the rotational distribution for different temperatures. Inset is the log scale plot. At room temperature, less than $1\%$ molecules are in $N=1$ state, while for 4K, it is enhanced to more than $10\%$. (b) The experimental data for different rotational populations. All data is normalized with with $N=0$ population. The lines are the theoretic curve for different temperature, and the best fitting shows the rotational temperature is $4.0(7)$ K.}
\end{figure}

\begin{figure}[t]
\includegraphics[width= 0.45\textwidth]{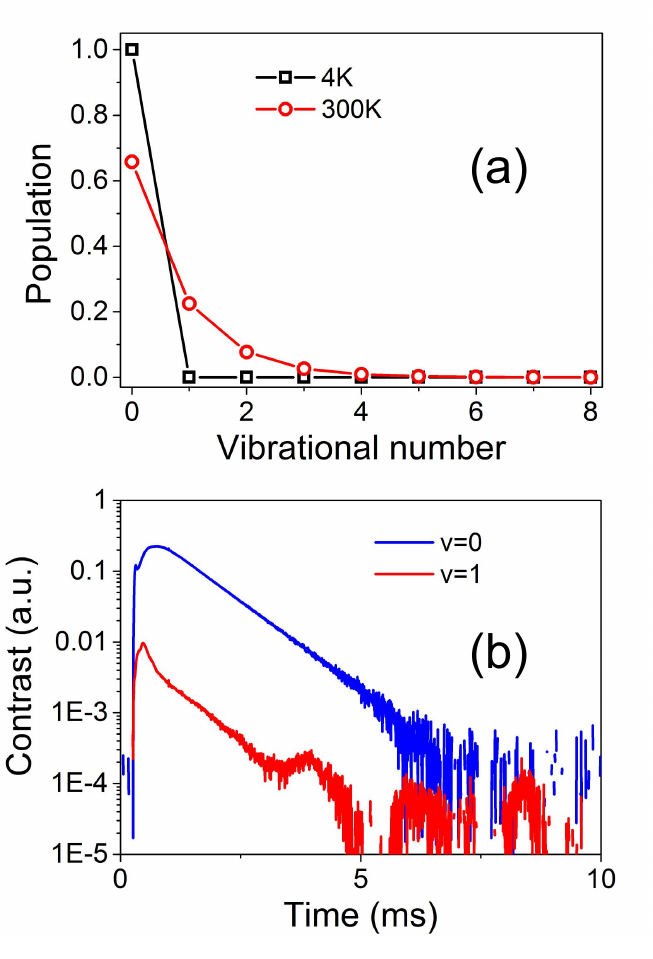}
\caption{\label{vibration}(Color online) The vibrational relaxation via buffer gas cooling. (a) The theoretic calculation of the vibrational distribution. At room temperature, a small fraction of molecules will populate higher-lying vibrational states. At $4$ K, almost all molecules populate the $v=0$ state. (b) The experimental absorption signal for $v=0$ and $v=1$ molecules from laser ablation. The population of $v=1$ is $10-100$ times smaller than $v=0$.}
\end{figure}

Another concern is the rotational and vibrational cooling via the buffer gas collisions. Generally, the energy difference between different vibrational states is about a few $1000$ K. Even at room temperature, only the ground vibrational state are largely populated, so it is not a big problem. However, for rotational states, the energy gaps are on the order of 1 K. At room temperature, molecules can populate to lots of different higher-lying rotational levels, as shown in Fig.\ref{rotational}(a). For the laser cooling scheme \cite{Chentao}, only a few low rotational states are involved. The small populated fractions lead to inefficient laser cooling.

According to the theoretic calculation of the rotational population at different temperatures in Fig.\ref{rotational}(a), by cooling the molecules from room temperature down to 4 K, the populated fractions for the lower-lying rotational levels can be enhanced by two orders of magnitude. And indeed, experimentally, after buffer-gas cooling of BaF, the populations at low rotational states get largely enhanced, as shown in Fig.\ref{rotational}(b). To measure the rotational fraction, we choose the transitions in the $P(N)$ branch, \emph{i.e.}, $\vert N,J\rangle \to \vert N'=N-1, J'=J-1\rangle$. The absorption strength reflects the population for the $N=1-8$ rotational states respectively. For $N=0$ state, we use the data from the $Q(0)$ measurement in Sec.\ref{subsectionC}. We plot the measured the populations at different rotational levels versus the rotational numbers, the best fits tell us the rotational temperature is $4.0\pm0.7$ K, which is consistent with the buffer gas temperature.

For the vibrational states, usually the buffer gas collision-induced quenching is far less efficient than that for the rotational states \cite{Campbell2009,Li2016,Egorov2001}. Theoretic calculation indicates (Fig.\ref{vibration}(a)) the population at $v=1$ for $T=4$ K is almost zero. But experimentally, we still observed the absorption signal from $v=1$ to $v'=0$. The signal is much weaker than the $v=0 \to v'=0$ transiton, as shown in Fig.\ref{vibration}(b), and it takes long time to average down the noise. We estimated the population of $v=1$ is about $10-100$ times smaller than that of $v=0$ state.

\section{\label{section7}Conclusion}

To conclude, we have created BaF molecules by laser ablation with a $532\ \text{nm}$ pulse laser, and buffer-gas cooled them with the 4 K helium gas. The vibrational and rotational degree of freedom has been effectively quenched. Almost all the molecules populate the $v=0$ level, and the rotational temperature is about 4 K. Large amount of molecules have been accumulated into the lower-lying rotational states, making further laser cooling experiments feasible. The collision cross section between BaF and helium gas is measured to be $1.4(7)\times10^{-14}\text{cm}^{-2}$, and it demonstrates that the BaF molecule is very suitable for buffer gas cooling. Meanwhile, the laser cooling relevant transition $|X^2\Sigma, v=0, N=1\rangle$ to $|A^2\Pi, v'=0, J'=1/2, +\rangle$ is observed with a high resolution. Our study paves the way for the laser cooling of BaF molecule.

\begin{acknowledgments}

We thank Jun Ye, Matt Hummon, Yong Xia, Liang Xu for helpful discussions. The authors acknowledge the support from the National Natural Science Foundation of China under grants 91636104, the Fundamental Research Funds for the Central Universities 2016QNA3007.

\end{acknowledgments}

\bibliography{BaF_spec}

\end{document}